\documentclass[11pt]{article}
\usepackage{iap2000,epsfig}

\def\be{\begin{equation}}
\def\ee{\end{equation}}
\def\bea{\begin{eqnarray}}
\def\eea{\end{eqnarray}}

\begin{document}
\vspace*{4cm}
\title{THE COMA CLUSTER LUMINOSITY FUNCTION FROM ULTRAVIOLET TO NEAR--INFRARED}

\author{S. ANDREON}
\address{Osservatorio Astronomico di Capodimonte,
via Moiariello 16, 80131 Napoli, Italy}
\author{J.-C. CUILLANDRE} 
\address{Canada--France--Hawaii Telescope,
PO Box 1597, Kamuela, 96743, Hawaii}
\author{R. PELL\'O}
\address{Observatoire Midi-Pyr\'en\'ees, LAT, UMR 5572,
14 Avenue E. Belin, F-31400 Toulouse, France} 

\maketitle

\abstracts{The Coma cluster luminosity function (LF) from ultraviolet ($2000
\AA $) to the near--infrared ($H$ band) is summarized. \hfill \break In the
$UV$ the LF is very steep, much steeper than in the optical. The steep
Coma $UV$ LF implies that faint and bright galaxies give similar contributions
to the total $UV$ flux and to the total metal production rate. The Coma $UV$
LF is dominated in number and luminosity by blue galaxies, which are often
faint in the optical. Therefore the Coma $UV$ LF is dominated by star forming
galaxies, not by massive and large galaxies.  \hfill \break The optical Coma
LF is relatively steep ($\alpha=-1.4$) over the 11 magnitudes
sampled, but its slope and shape depend on considered filter and magnitude.
We found a clear steeping of the FL going from $B$ to $R$ bands, indicative of
the presence of a large number of red dwarfs, as faint as three bright
globular clusters. Furthermore, using {\it Hubble Space Telescope} images, we
discover that blends of globular clusters, not resolved in individual
components due to seeing, look like dwarf galaxies when observed from the
ground and are numerous and bright. The existence of these fake extended
sources increases the steepness of the LF at faint magnitudes, if
not deal on. This concern affects previous deep probing of the luminosity
function, but not the present work. \hfill \break
The near--infrared LF was computed on a near--infrared selected sample of
galaxies which photometry is complete down to the typical dwarf ($M^* +5$)
luminosity. The Coma LF can be described by a Schechter function with
intermediate slope ($\alpha\sim-1.3$), plus a dip at $M_H\sim-22$ mag. The
shape of the Coma LF in $H$ band is quite similar to the one found in the $B$
band.  The similarity of the LF in the optical and $H$ bands implies that in
the central region of Coma there is no new population of galaxies which is too
faint to be observed in the optical band (because dust enshrouded, for
instance), down to the magnitudes of dwarfs. The exponential cut of the LF at
the bright end is in good agreement with the one derived from shallower
near--infrared samples of galaxies, both in clusters and in the field. The
faint end of the LF, reaching $M_H\sim-19$ mag (roughly $M_B\sim -15$), is
steep, but less than previously suggested from shallower near--infrared
observations of an adjacent region in the Coma cluster. 
}

\section{Introduction}

Among nearby clusters of galaxies, Coma ($v\sim7000$ km s$^{-1}$) 
is one of the richest ($R=2$) ones. At a first glance, it looks relaxed and
virialized in both the optical and X-ray passbands. For this reason it
was designed by Sarazin (1986) and Jones \& Forman (1984) as the
prototype of this class of clusters. The optical structure and photometry
at many wavelengths, velocity field, and X-ray appearance of the cluster
(see the references listed in Andreon 1996) suggest the existence of
substructures. Since these phenomena are also observed in many other
clusters (Salvador-Sol\'e, Sanrom\`a, \& Gonz\'ales-Casado 1993), the Coma
cluster appears typical also in this respect. Therefore, the Coma cluster,
is an ideal target for galaxy studies.

The luminosity function (LF) represents the zero--order statistics of galaxy
samples and gives the relative number of galaxies as a function of the
magnitude. Almost every quantity is, therefore, ``weighted" by the LF,
including obvious quantities, such as the galaxy color distribution, and also
less obvious ones, such as correlations (see, for example, the discussion on
the impact of magnitude limits in the size--luminosity relation by Simard et
al. 1999). When the sample is not complete in volume a further ``weight"
should be added: the selection function. Thus, an accurate knowledge of the LF
is important for galaxy studies.

Since different wavelengths carry different informations,
the determination of the LF at different wavelenghts is interesting.

In the near--infrared, k corrections are relatively small and well known, thus
allowing to observe and to fairly compare galaxies at different redshifts, up to
high redshift values. In particular, k corrections are almost independent
from the spectral type of galaxies, in such a way that statistics on a
population of galaxies are less affected by changes of the morphological
composition induced by differential corrections from type to type.
Furthermore, galaxies that undergo a starburst are not selected
preferentially, as instead happens in the optical, and therefore a sample
selection in the near--infrared is less biased by episodic events of star
formation. Finally, the near--infrared luminosity is a good tracer of the
stellar mass.

In both the single stellar population and continuous star formation scenarios,
the $UV$ luminosity of late--type galaxies appears to be largely dominated by
young massive stars, thus implying a direct link between $UV$ luminosities and
star formation rates (e.g. Buzzoni 1989). For all but the very old stellar
populations, the $UV$ traces mainly the emission from young stars (see for
instance Donas et al. 1984; Buat et al. 1989), having maximum main sequence
lifetime of a few $10^8$ years. Therefore, for star forming galaxies the $UV$
is a direct measure of the present epoch star formation rate. 

The optical luminosity of galaxies provides instead a weighted average of the
past to present star formation rate. This is also the waveband observationally
easier, allowing us to reach resolution, coverage and dept difficult to
achieve in the $UV$ or the near--infrared bands.

This paper summarizes our recent results on the Coma cluster luminosity 
function at various wavelenghts. LFs are computed from 
recent photometric surveys with the FOCA balloon in the $UV$, CFH12K at
CFHT in the visible, and Moicam at Pic du Pidi in the near infrared.

\section{Summary of photometric data}

Coma was observed in the $UV$ by Donas, Milliard, \& Laget (1991 \& 1995) 
with a panoramic detector (FOCA, see Milliard et al. 1991). Images have been
taken with a filter centered at 2000 \AA \ with a bandwidth of 150 \AA \ and
it is complete down to $UV\sim17-17.5$ mag and 70\% complete in the range
$17.5<UV<18$ mag. Almost the whole cluster (a circle of 1 deg of radius) 
has been observed.

Multicolor Coma cluster observations have been taken during the CFH12K
(Cuillandre et al. 2000) first light at the Canada--France--Hawaii telescope
prime focus in photometric conditions. CFH12K is a 12K $\times$ 8K CCD mosaic
camera, whose field of view is 42 $\times$ 28 arcmin$^2$ with a pixel size of
0.206 arcsec. Figure 1 shows our $V$ image of the studied portion of the Coma 
cluster. After discarding areas noisier
than average (gaps between CCDs, borders, regions near bright stars and large
galaxies, etc.), the usable area for the Coma cluster is 0.29 square degree in
$V$ and $R$ and 0.20 degree sq. in $B$, i.e. $\sim 1000$ and $\sim 700$
arcmin$^2$, respectively.

The near--infrared Coma LF has been computed from the photometry presented in
Andreon, Pell\'o, Davoust et al. (2000), to which we defer for details. In
summary, a $\sim 20 \times 24$ arcmin region of the Coma cluster, located
$\sim 15$ arcmin from the centre, have been imaged with the Moicam camera at
the 2.0m Bernard Lyot telescope at Pic du Midi. Images were taken in the $H$
band under moderate to good seeing conditions (i.e. $1<FWHM<1.5$ arcsec), with
average exposure time of $\sim300$ sec. About 300 objects have been detected
and classified by Sextractor version 2 (Bertin \& Arnouts 1996) in the best
exposed part of our mosaic ($\sim380$ arcmin$^2$).

The $H$ band image of studied portion of the Coma cluster is show in
Figure 2. It is co--centered to the optical image shown in Figure 1.

\begin{figure}
\caption{The image shows the whole CFH12K field of view 
$V$ image of the studied field. North is up and east is to the left. 
The field of view is $42\times28$ arcmin$^2$, i.e. $1.2\times0.8$ Mpc$^2$
at the Coma distance (assuming $H_{0}=68$ km s$^{-1}$ Mpc$^{-1}$). 
Regions with lower quality than average are not considered
(such as the bottom right CCD).} 
\end{figure}

\begin{figure}
\caption{
$H$-band mosaic of the region under investigation. 
Faint objects have $H\sim16$ in this heavly rebinned
and compressed (for display purpose) image.
The field is $\sim 20 \times 24$ arcmin. North is up 
and East is left. The two dominant galaxies of the Coma cluster are located 
near the South-West corner. Compare with the optical appearance of
the same field (Figure 1)}
\end{figure}

\section{The $UV$ Coma LF}

\begin{figure}
\centerline{\psfig{figure=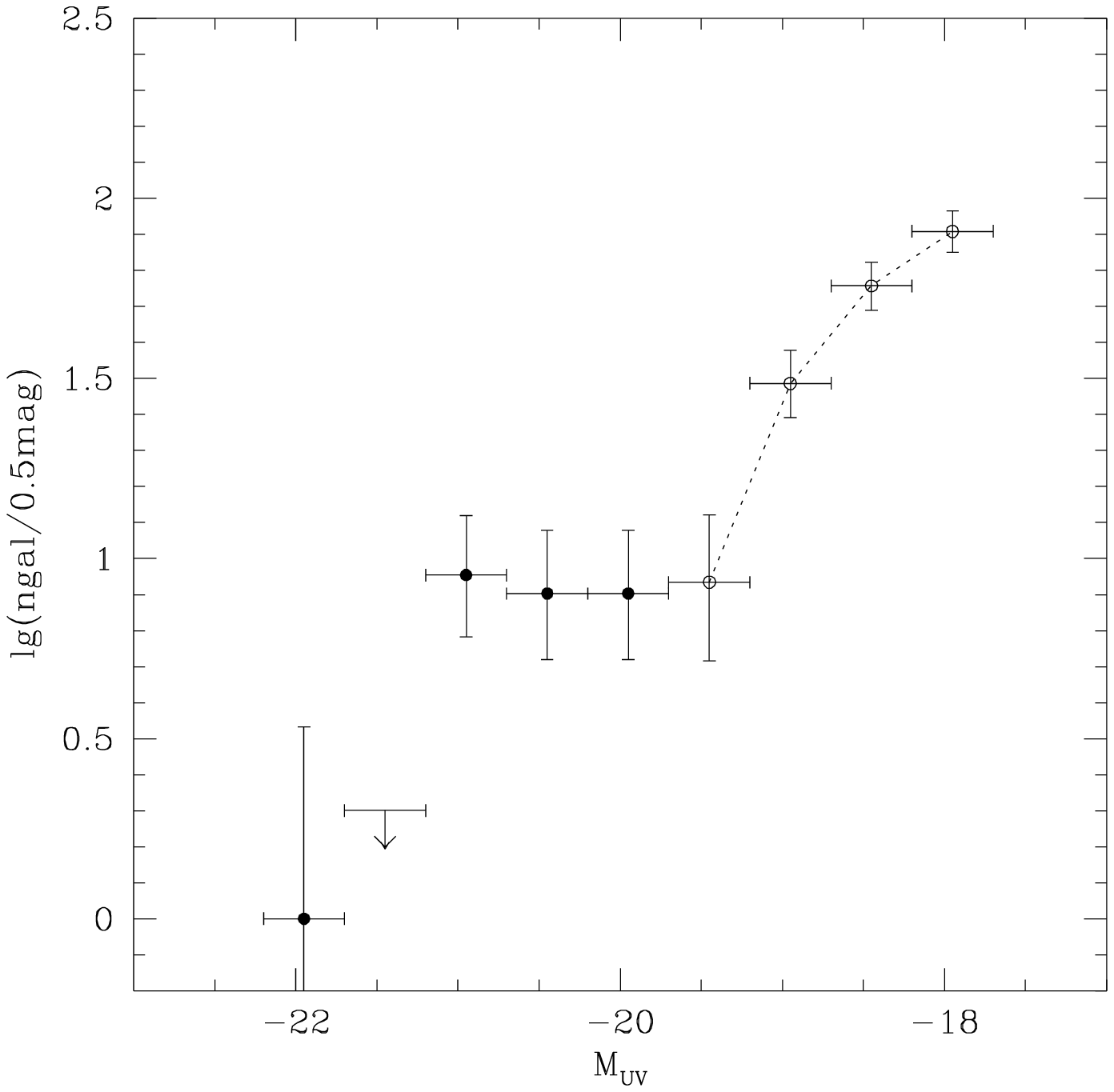,width=8truecm}
\psfig{figure=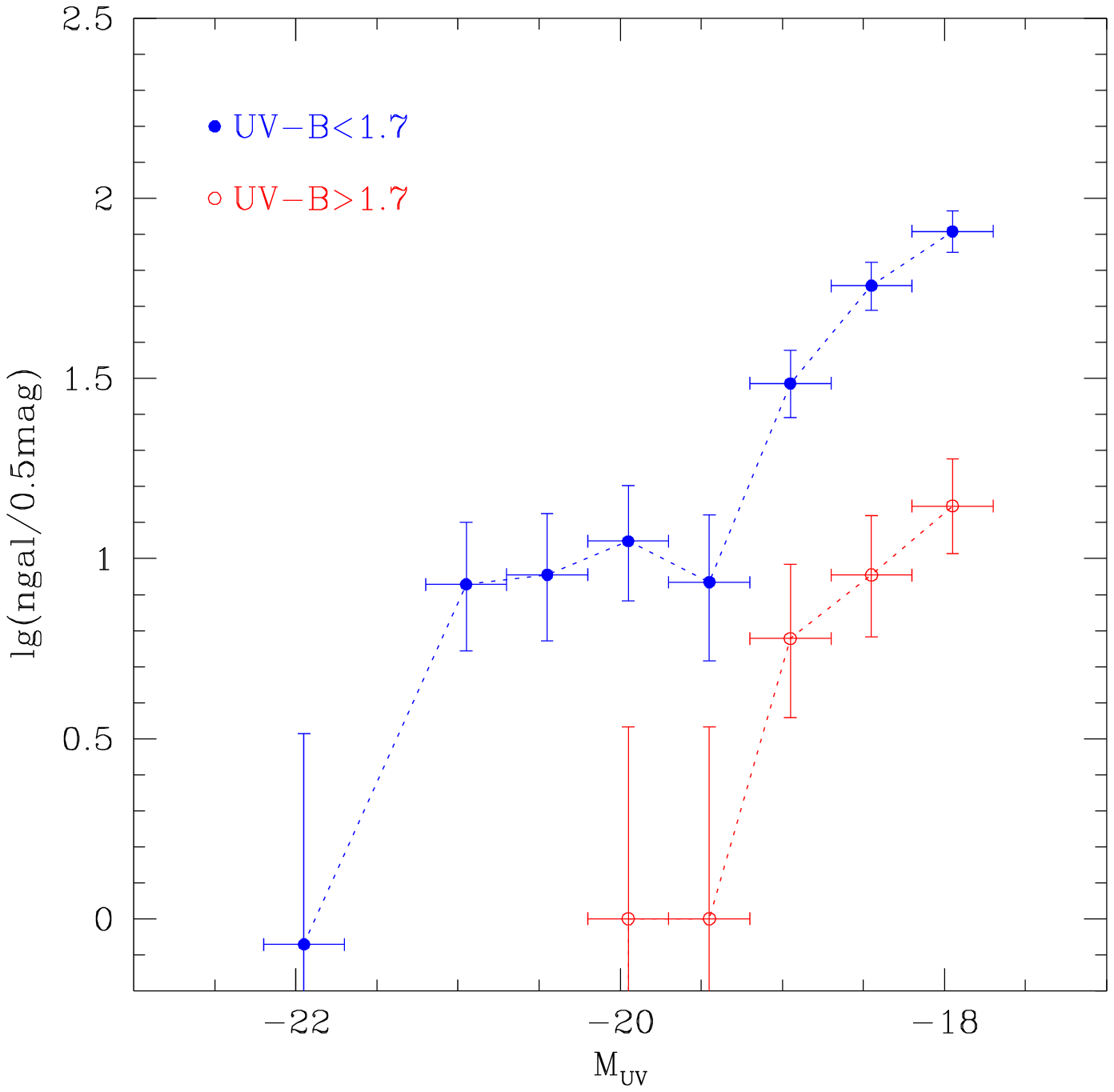,width=8truecm}}
\caption{{\it Left}: 
The $UV$ LF of Coma cluster galaxies.
Open points are computed using redshift for assigning
the cluster membership, close points use a statistical subtraction of the
interlopers. {\it Right}: 
Bivariate LF of Coma galaxies. Close points
refer to blue ($UV-b<1.7$) galaxies, open points refer to red ($UV-b>1.7$)
galaxies. In this figure we adopt $H_{0}=50$ km s$^{-1}$ Mpc$^{-1}$.
}
\end{figure}

The Coma UV LF is computed in a quite complex way that use both statistical
methods and available redshift surveys in the Coma cluster direction (see
Andreon 1999 for details). Three improvements apply since the Donas, Milliard,
\& Laget (1991) paper: the essential galaxy background counts are now available
for the LF computation, the number of known redshifts  is increased by 60\% and
a more elaborate statistical treatment is adopted. Figure 3 shows the UV LF. It
is the first ever derived for a cluster. Error bars are large, and only the
rough shape of the LF can be sketched. 

The Coma $UV$ LF is well described by a power law (or alternatively by a
Schechter function with a characteristic magnitude $M^*_{UV}$ much brighter
than the brightest galaxy): a Kolmogorov-Smirnov test could not reject at more
than 20\% confidence level the null hypothesis that the data are extracted
from the best fit (whereas we need a 68\% confidence level to exclude the
model at $1\sigma$). The best slope is $\alpha=0.40\pm0.03$. In terms of the
slope of the Schechter (1976) function $\alpha_S$, this values is $-2.0$. It
needs to be stressed, however, that the computed slope of the LF (and in
particular its error) depends on the backgound counts, which are quite
uncertain.

The steep Coma $UV$ LF implies that faint and bright galaxies give similar
contributions to the total $UV$ flux, and that the total $UV$ flux has not yet
converged 4 magnitude fainter than the brightest galaxy (or, equivalently, 
at $M_3+3$). Therefore, in order to derive the total luminosity and
hence the metal production rate, it is very important to measure the LF down
to faint magnitude limits. 

The right panel of Figure 3 shows the bi--variate color--luminosity function,
i.e. the LF of red ($UV-b>1.7$ mag) and blue ($UV-b<1.7$ mag) galaxies. The
bulk of the $UV$ emission comes from blue galaxies while all red galaxies have
$M_{UV}>-20$ mag. Therefore, since blue galaxies dominate the $UV$ LF both in
number and luminosity, the Coma $UV$ LF is dominated by star forming galaxies
and not by massive galaxies, which are often faint in $UV$.  From previous
morphological studies (Andreon 1996) it turns out that Coma red galaxies in our
sample are ellipticals or lenticulars. The fact that early--type galaxies
contribute little to the $UV$ LF may be explained as a consequence of the fact
that these systems have a low recent star formation histories. Please note that
in the optical, the LF is dominated at the extreme bright end by the
early--type (i.e. red) galaxies (Bingelli, Sandage \& Tammann 1988, Andreon
1998), and not by blue ones as it is in $UV$.

\begin{figure}
\centerline{%
}
\caption{Residual $R$ image after having subtracted a model of
IC\,4051 from the original image. The left panel shows the ground
image, while the right one displays the {\it HST} one. 
The galaxy is much larger than the
field of view of this cutout, whose angular size is 68 arcsec. 
In the left panel, the square marks IC\,4051 center, and circles the true galaxies
in the field, as confirmed by the {\it HST} image. The remaining 
objects in the left panel, most of which are extended sources and 
are brighter than the 
completeness magnitude, are unresolved blends of GCs.}
\end{figure}

\section{The optical Coma LF}

The LF (or, equivalently, the relative space density distribution of galaxies
of each luminosity) is computed as the difference between galaxy counts in the
Coma and in the control field directions (for an introduction on the method,
see, e.g., Oemler 1974; Zwicky 1957). All technicalities, as well as results
concerning the optical Coma LF, are described in Andreon \& Cuillandre (2000).

\begin{figure*}
\hbox{
\epsfysize=6.5cm
\epsfbox[45 195 560 540]{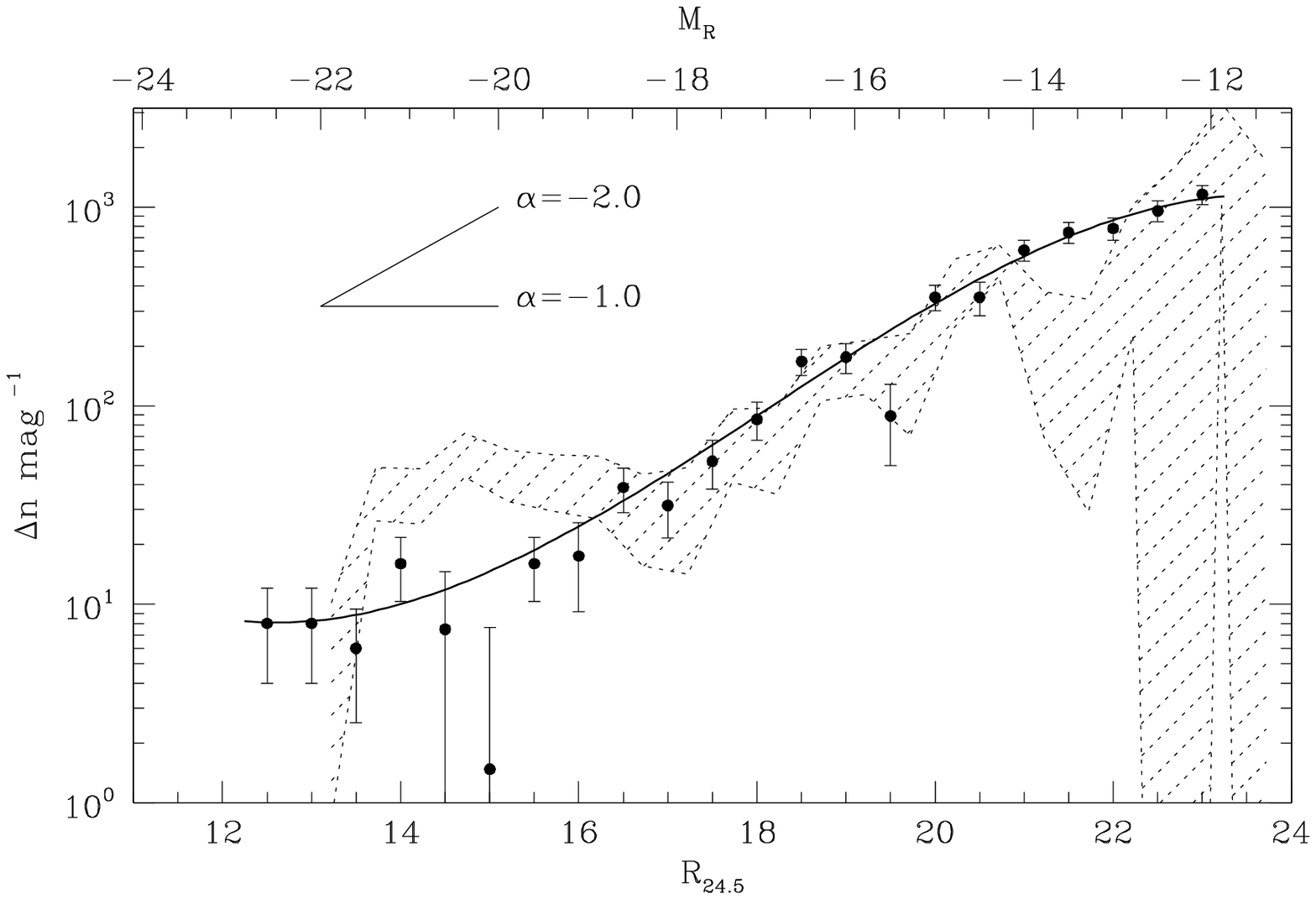}
\hfill\null}
\hbox{
\epsfysize=6.5cm
\epsfbox[45 195 560 540]{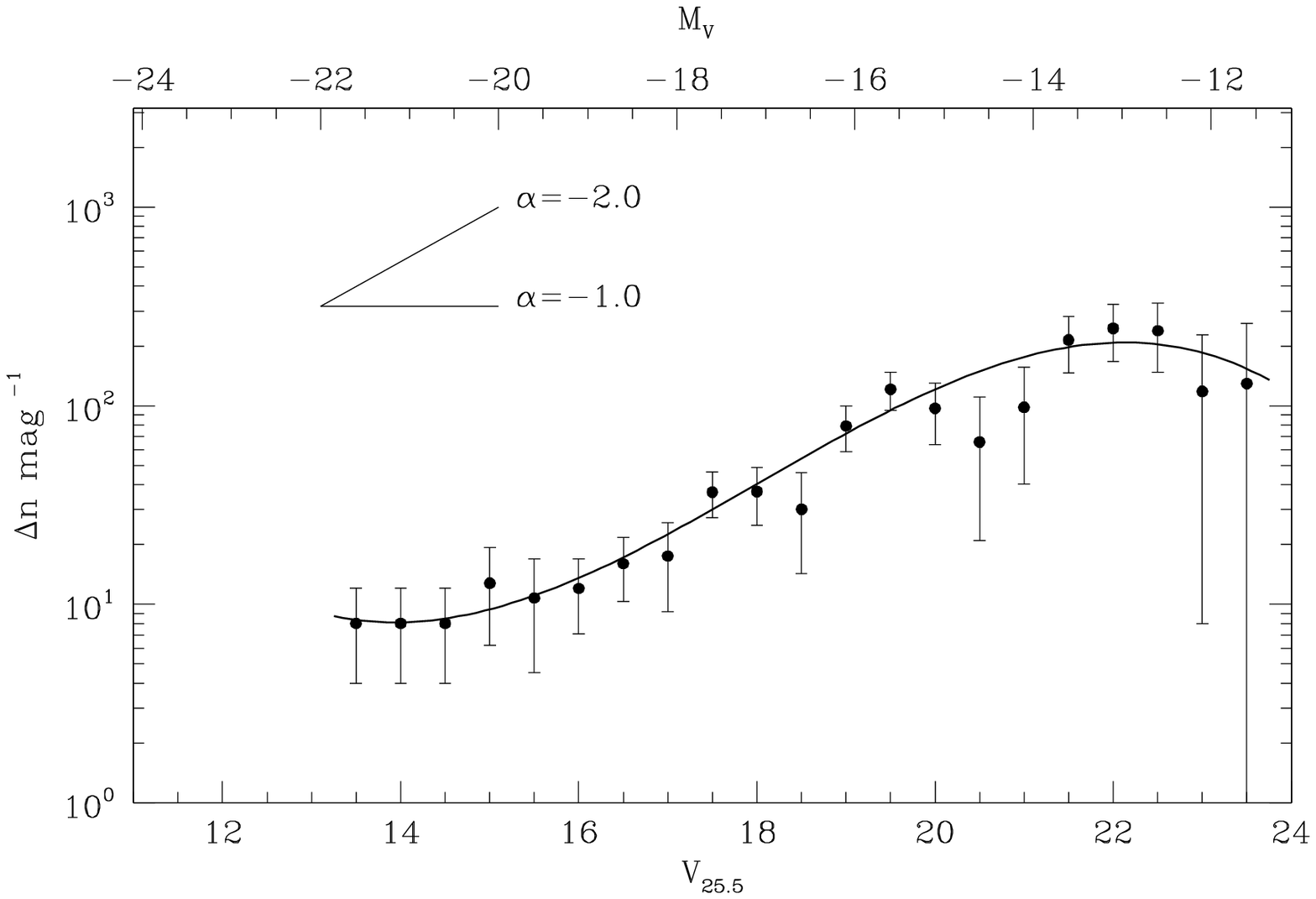}
\hfill\null}
\hbox{
\epsfysize=6.5cm
\epsfbox[45 195 560 540]{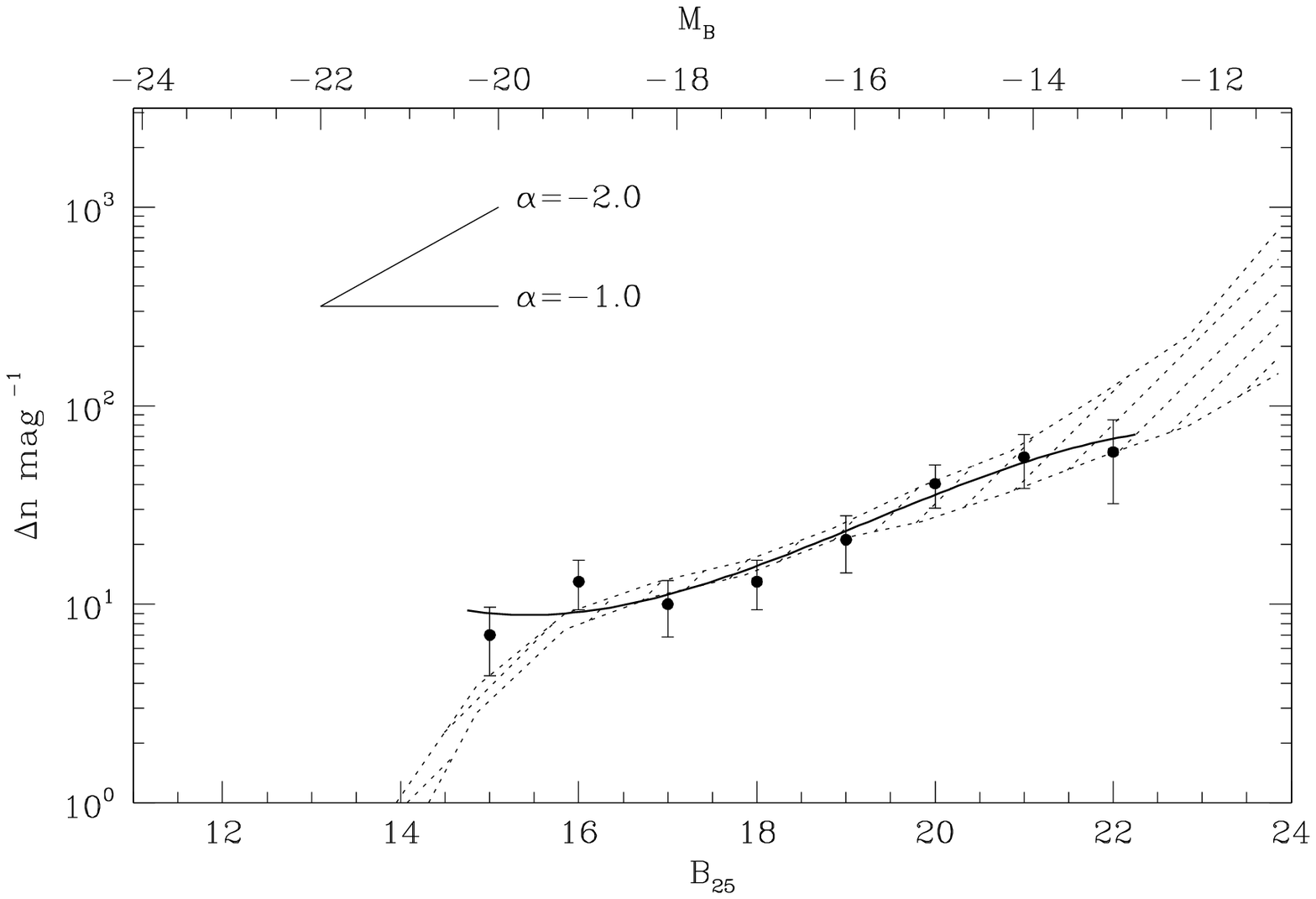}
\hfill\null}
\caption{LF of Coma cluster in the $R$, $V$ and $B$ bands
(close points). Both apparent and absolute magnitudes are shown on the
abscissa. The upper abscissa assumes 
$H_{0}=68$ km s$^{-1}$ Mpc$^{-1}$.
The thick curve is the best fit with a 3 degree 
power--law to the data. In the $R$ (top) panel, the hashed region delimit
the LF determined by Trentham (1998a). 
In the $B$ (bottom) panel, the hashed region
delimit the composite $B$ LF, averaged over almost all literature
ones (from Trentham 1998b).}
\end{figure*}

Globular clusters (GCs hereafter) are a source of concern when dealing with
faint probings of the LF, because their blends contaminate galaxies counts. We
discover that this possibility actually arises by comparing archive {\it Hubble
Space Telescope} images of IC\,4051 (Baum et al. 1997), which is an early--type
galaxy of the Coma cluster in our field of view, to our ground image. Figure 4
shows the residual image of IC\,4051, as seen in our ground image, after
subtracting a model of the galaxy (obtained by fitting its isophotes). The only
actual galaxies are marked in the left panel by circles. Notice that only one
faint object is circled, i.e. it is an actual galaxy. All the other objects are
blends of a few GCs, unblended at the {\it HST} resolution. Most of them are
brighter than our (and other deep probing of the LF) completeness limit, and
therefore would be counted as galaxies in the LF. Other GCs blends are present
near NGC\,4481 ({\it HST} images are published in Baum et al. 1995), another
bright Coma elliptical in our field of view. The density of these extended
(i.e. counted as galaxies) objects can be huge near bright galaxies. Near
IC\,4051, we found $3 \times 10^6$ candidate GCs blends per mag per square
degree at $R\sim25.5$, which is 40 times larger than background galaxies of the
same luminosity. The brightest of them has $R=20$ mag, i.e. there are GC
blends $\sim7.5$ mag brighter that the GC turnoff (directly measured by Baum et
al. (1997) for this galaxy). Therefore, GC blends do not affect only GCs
typical magnitudes ($V\sim27$ mag), but also much brighter counts and are
pernicious because they are extended sources. Previous works studying the
deepest part of the galaxy LF are likely affected by GC blends at faint
magnitudes. Thus, the points found in literature at $M>\sim-15$ mag should be
regarded with caution as long as the area surveyed is comparable (or smaller)
than that occupied by bright galaxies. Because of this potential source of
error, we generously mask a few bright galaxies with a large GC population and
a halo. Residual unflagged contamination is diluted by the very large field of
view of our images.

Figure 5 shows (filled points) the Coma cluster LF down to
$R\sim 23.25, V\sim 23.75$ and $B\sim22.5$ mag. Notice the large number of
galaxies per magnitude bin in our $R$ and $V$ LFs and the absolute faintness
of studied galaxies ($M_R\sim-11.75, V\sim-11.25$, and $M_B\sim -13$ mag),
whose luminosity exceed the tip of the globular cluster LF ($M_V\sim-10$ mag)
by less than a factor 3 in flux (in the deepest bands). The LF extends over an
11 magnitude range and it is one of the deepest every confidently measured
from CCD photometry. 

In the $R$ band, the LF shape is not well described by Schechter (1976) law, 
because its best fit has $\chi^2\sim37$ for 18 degrees of freedom. A
function with more free parameter better describes the data, as almost always
happens when the data quality is good. The best fit with a 3$^{rd}$ order
power--law (i.e. with one more free parameter) is overplotted in Figure 5
(smooth curves). The reduced $\chi_{\nu}^2$ is $\sim 1$, suggesting a good
fit. 

The LFs in the three filters present both similarities and differences. The
LFs seem truncated at the bright end ($R=12$, $V=13.5$, $B=15$ mag), then they
are fairly flat at intermediate luminosities ($B<18, V<16$ and $R<16$ mag). At
fainter magnitudes, the LFs are steep in $R$ and $V$, and with a much
shallower slope in $B$. Of course, the exact slope depends on the considered
magnitude and filter. As a guideline, the typical slope range from $-1.25$ in
$B$ to $-1.6$ in $R$. There is, therefore, indirect evidence that most of the
faint galaxies in the Coma cluster are red. This indirect evidence can be
tested with the data we have, because the individula galaxy color can be 
computed. This investigation will be included in a future paper, where we 
shall present the LF bi--variate in color, i.e. the luminosity distribution of
galaxies of a given color.

In $R$ and $B$ no turn--off is seen in the LF at faint magnitudes: the minimal
luminosity of dynamically linked object (galaxies) can be as low as the
luminosity of just three bright globular clusters. Furthermore, these very
faint galaxies dominate, by number, the galaxy LF. In $V$ there is an hint of
flattening of the LF at faint magnitudes, however at these faint magnitudes
errors are quite large, and therefore any definitive conclusion can be drawn.

The shaded regions in Figure 5 delimit the best previous determinations of the
LF. In the $R$ band, the shaded region is the LF of the ``deepest and most
detailed survey covering [{\it omissis}] a large area" by Trentham (1998a).  In
the $B$ band, Trentham (1998b) summarizes our present knowledge on the LF by
computing the composite cluster LF, averaging over almost all literature LFs
based on wide field deep images, including seminal papers, such as Sandage,
Bingelli \& Tammann (1985), for example. The shaded region in the bottom panel
shows his result. Our determination is at least of comparable quality with
respect to the best previous efforts (see, in particular, the large errorbar of
the  Trentham (1998a) $R$ band LF at faint magnitudes).

To summarize, we compute one of the deepest LF in three bands, over a very
large magnitude range (up to 11 mag), with good statistics and our results
agree with previous LF determinations on the common magnitude range. We found a
clear steeping of the FL from $B$ to $R$, indicative of the presence of a large
population of red dwarfs in Coma. We also found no turnoff in the Coma LF, i.e.
galaxies can be as faint as 3 bright GC, and these galaxies outnumber much
brighter galaxies. The derived LFs, in $B$, $V$ and $R$ have the following
pluses (discussed in Andreon \& Cuillandre 2000): 1) we have a complete census
(in the explored region) of low surface brightness galaxies with central
surface brightness almost as low as the faintest so far cataloged galaxies; 2)
the explored area is among the largest ever sampled with CCDs at comparable
depth for any cluster of galaxies; 3) the error budget includes all source of
errors known to date; 4) our derivation of the LF does not discard a priori
compact galaxies, as most previous LF determinations (this last point
is discussed in Andreon \& Cuillandre 2000). Furthermore, using {\it
Hubble Space Telescope} images, we discover that blends of globular clusters,
not resolved in individual components due to seeing, look like dwarf galaxies
when observed from the ground and are numerous and bright. The existence of
these fake extended sources increases the steepness of LF at
faint magnitudes, if not deal on. This concern affects previous deep probing of
the LF, but not the present work.

\section{The near--infrared Coma LF}

\begin{figure}
\centerline{\psfig{figure=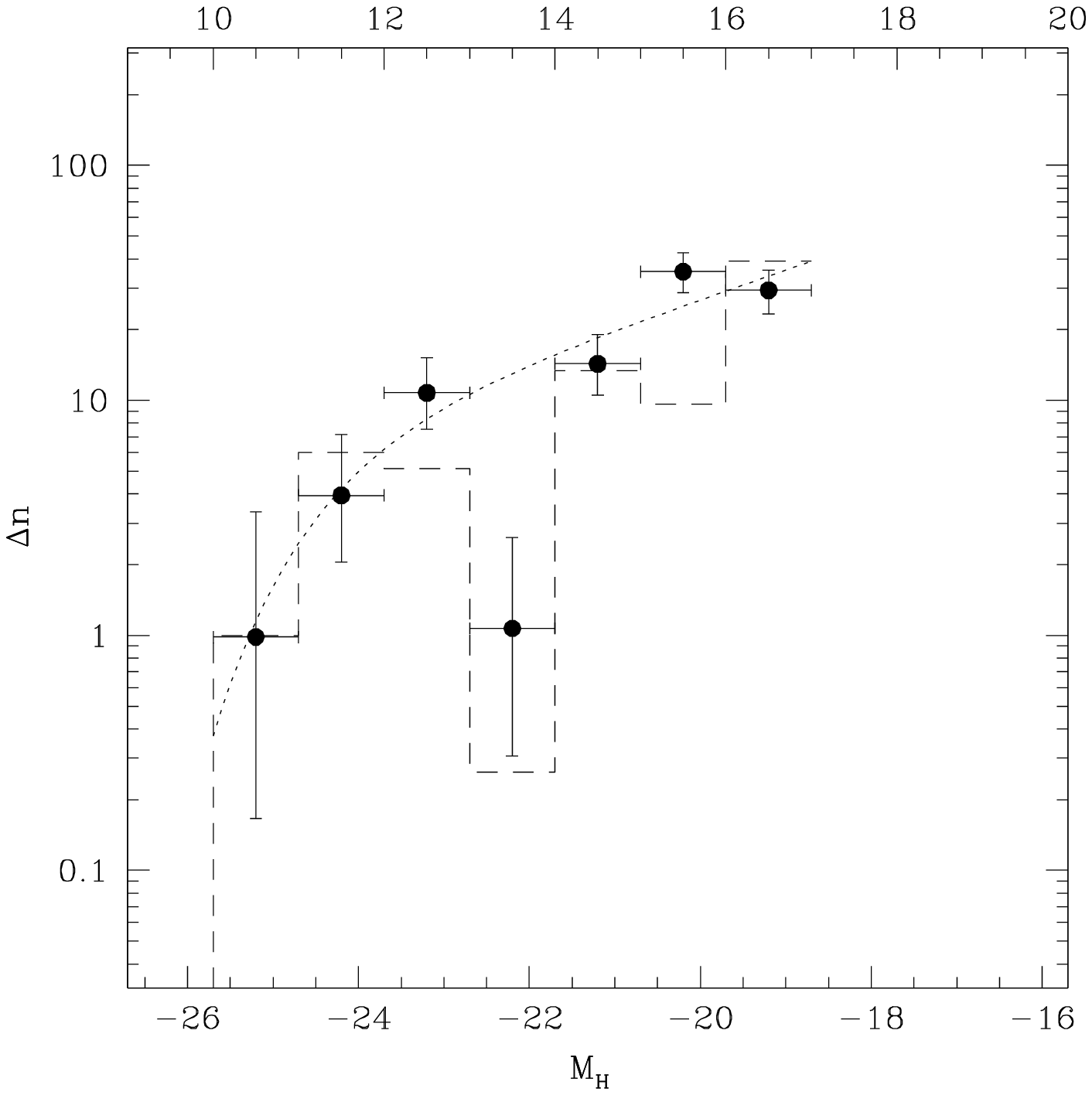,width=8truecm}
\psfig{figure=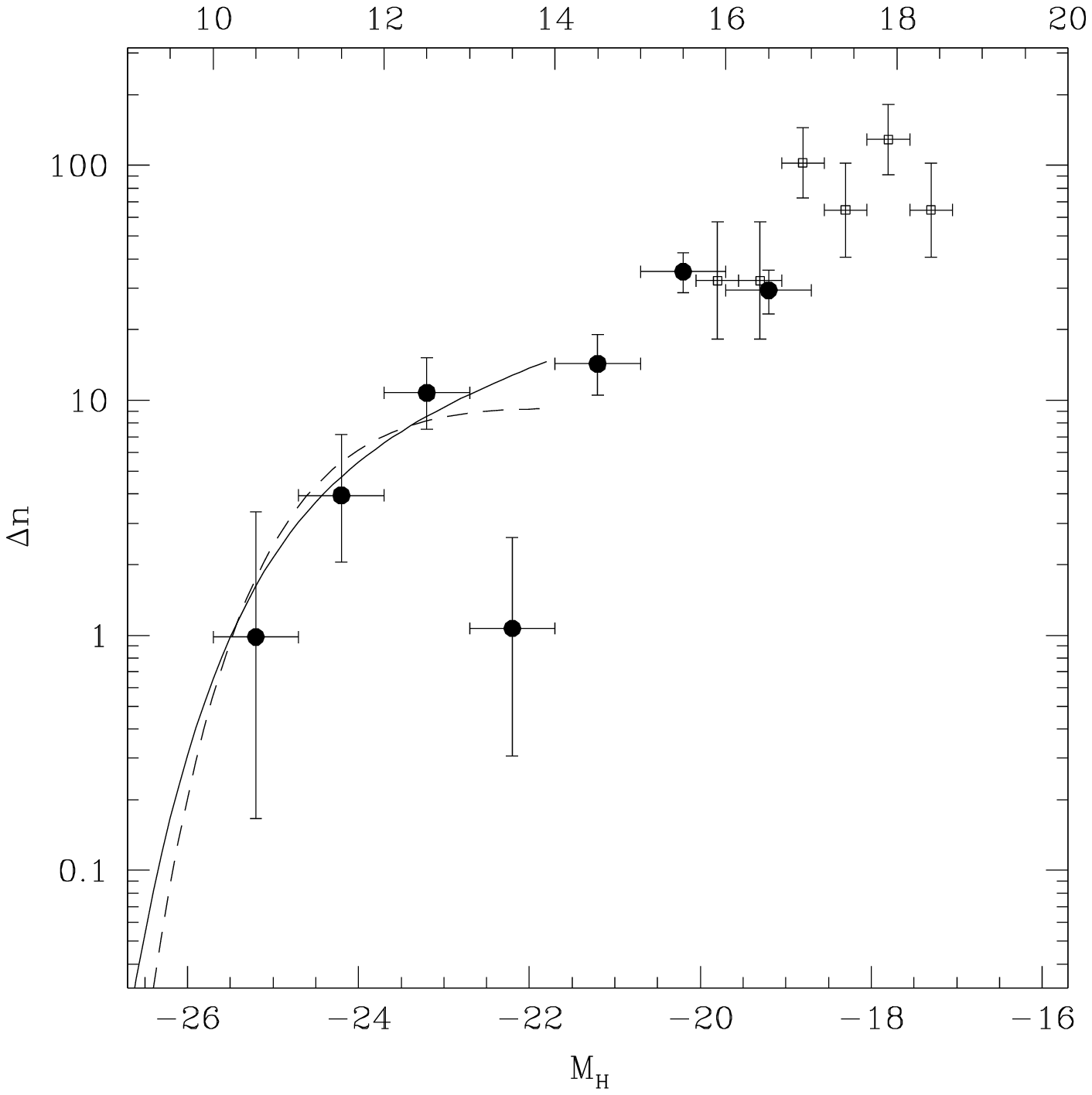,width=8truecm}}
\caption{{\it Left}: Coma LF in the $H$ band, as computed from the present data alone
(closed dots). $\Delta n$ is the number of Coma galaxies
per magnitude bin in the studied field.
The dashed histogram is the $b$ Coma LF, shifted by a pseudo-color
$b-H=3.5$ mag. The dotted line is the best fit by a Schechter function, once
the dip point is flagged. {\it Right}:
Various determinations of the near--infrared LF. Our own data (solid dots) and
Mobasher \& Trentham (1998) data (open squares) are shown, after normalization 
of the LF in common bins. Local field LFs are also shown:
Gardner et al. (1997) (dashed line) and Szokoly et al. (1998) (solid line).
The field LF has been vertically shifted to reproduce 
the Coma LF in the three brightest bins.
The upper abscissa scale in both panels shows the apparent $H$
magnitude, and the lower one gives the corresponding absolute $H$ magnitude,
adopting $H_{0}=50$ km s$^{-1}$ Mpc$^{-1}$.} 
\end{figure}

The Coma cluster LF is computed statistically as we did in the optical, and
fully presented in Andreon \& Pell\'o (2000).  In particular, we
take into account that an expected and important source of error in
the LF determination is the background variance from field to field.

The near--infrared Coma LF is presented in Figure 6. It is characterized by
a bright end (at $M_H\sim-25$ mag), a part increasing gently down to
$M_H\sim-18.5$ mag, and an ``outlier" point at $M_H=-22.2$ mag, which
produces a dip in the Coma LF. The LF displayed in Figure 6 is the
deepest ever measured in any near--infrared band for any type of
environment for a near--infrared selected sample, reaching 
$\sim M_H^*+6$, well in the dwarf regime.

In the optical, the Coma cluster exhibits a similar shape (Godwin \& Peach
1977; Secker \& Harris 1996). Using Godwin, Metcalfe \& Peach (1983) data, we
computed the $b$ (a photographic $J$-like blue filter) Coma LF in almost
exactly the same area surveyed in the $H$ band\footnote{The optical LF
computed in this way seems differ to those computed in $B$ from CCD data.
The cause of this possible difference is under investigation}. 
Then, because the $H$ and $b$
band magnitudes are available for all galaxies in this area, we have computed
a mean $b-H$ pseudo-color (magnitudes are not measured at the same aperture).
No normalization in $\Delta n$ has been applied. The result of this exercise
is shown in Figure 6 as a dashed-line histogram. The two LFs are remarkably
similar: there is a close agreement between the cut at $M_H=-25$ mag, the dip
location ($M_H=-22.2$ mag), the dip amplitude and the increase observed at
fainter magnitudes (closed dots) and expected from the $b$ LF (dashed
histogram). Thus, the shape and the amplitude of the Coma LF seems not to be
strongly depend on the wavelength when we compare the results in $b$ and in
$H$ bands.

The strong similarity of the optical and near--infrared LFs implies that in the
near--infrared there is no new population of galaxies which disappears in the
optical band (because dust obscured, for example), down to the magnitude of
dwarfs. Furthermore, if the $H$ band LF traces the galaxy mass function in
this cluster, the same holds true for the blue LF. This result has been
obtained in a particular region of Coma, and could not be generalized without
further measures in other environments (field, groups, etc.). 

The existence of a dip in the near--infrared LF, which turn out to be not a
statistical fluctuation of the smooth LF at the 99.95 \% confidence level,
allows to discard the hypothesis, suggested by Lobo et al. (1997) that the
galaxies brighter than the dip were subjected to a recent episode of star
formation induced by the hostile Coma environment, which have made them
brighter. In that case, in fact, the dip should be absent or at least highly
attenuated in $H$, because the near--infrared luminosity traces the galaxy
mass and it is less affected by the short timescale starbursts that make a
few galaxies brighter than the magnitude of the dip. 

The exponential cut of the LF at the bright end is in good agreement with the
one derived from shallower near--infrared samples of galaxies, both in
clusters and in the field (see the right panel of Figure 6). This fact is
suggestive of a similarity of the tip of the mass function of galaxies,
irrespective of the environment where they are found. The dip at $M_H\sim-22$
mag is instead unique among all the so far measured near--infrared LF,
although several published observations are not deep enough or spanning a
suitable wide field to distinctly detect this feature. The faint end of the
LF, reaching $M_H\sim-19$ mag (roughly $M_B\sim -15$), is steep, but less than
previously suggested from shallower near--infrared observations of an adjacent
region in the Coma cluster (de Propris et al. 1998).

The overall slope of the Coma LF is intermediate ($\alpha\sim-1.3$). The slope
is measured down to the dwarfs regime: we reach $M_B\sim-14.5$ using own our
data alone and even fainter magnitudes ($M_H\sim-17$, roughly equivalent to
$M_B\sim-13$) when including Mobasher \& Trentham (1998) data and under their
assumptions. 

\section{Short summary and future perspectives}

In the previous sections we presented the zero order statistics, i.e. the
LF, of galaxies in the Coma cluster, as derived by studying {\it
independently} our ultraviolet, optical, and near--infrared images.

The {\it combined} use of the various datasets offers the unique opportunity of
study several other quantities, such as the LF bivariate in color (any
combination of ultraviolet, optical and near--infrared color) or the color
distribution at each magnitude, because all observations share a 0.3 deg sq. 
common field of view. The multicolor approach should allow to have a better understanding of the
galaxies properties.

The same data allow also the determination of high order moments of the
galaxy distribution, such as the luminosity density ($L \times LF$). The
brightness function, i.e. the number of galaxies of a given central
brightness, is also within the reach of the optical and near--infrared data,
allowing, for example, to determine whether it is bimodal (Tully \& Verheijen
1997) or not.

 \section*{Acknowledgments}

Based on observations collected with the {\it T\'elescope Bernard Lyot}, at
the Pic du Midi Observatory,  at the Canada--France--Hawaii Telescope and in
part at the {\it Hubble Space Telescope}. We thanks Florence Durret and Daniel
Gerbal for the nice and interesting meeting they organized.

\end{document}